\documentclass[aps,pre,floatfix,amsmath,amssymb,superscriptaddress,12pt]{revtex4-1}

\usepackage{amsthm}
\usepackage{float}
\usepackage{verbatim}
\usepackage{url}
\usepackage{graphicx}

\newcommand{\beq}{\begin{equation}}
\newcommand{\eeq}{\end{equation}}
\newcommand{\beqa}{\begin{eqnarray}}
\newcommand{\eeqa}{\end{eqnarray}}

\usepackage{xcolor}

\begin{document}

\title{COVID-19 in air suspensions}
\author{Daniel A. Stariolo}
\affiliation{Departamento de F\'{\i}sica,
Universidade Federal Fluminense and
National Institute of Science and Technology for Complex Systems,\\
Av. Litor\^anea s/n, Campus Praia Vermelha,
24210-346 Niter\'oi, RJ, Brazil}

\date{\today}

\begin{abstract}
We analyze the stability of virus-carrying particles in air at equilibrium after the
dissipation of the initial turbulent process produced by sneezing, coughing, breathing
or speaking. Because the viruses are expelled mainly attached to small liquid droplets,
with diverse sizes and weights, and the external environmental conditions can also
be diverse, the subsequent motion spanes different spatial and temporal scales. We
analize the validity of two simple models in this context. For
droplet sizes larger than $100\,\mu m$, computing the time of decay to the ground and
the distance travelled within a simple free fall model with empirical data extracted
from the literature, we obtain distances in the range between 1 to 3 meters from the
emitter, with a falling time of less than $1\, s$, similar to known recommendations for
safe social distancing. For droplet sizes less than $100\,\mu m$ a model of motion
in a viscous medium predicts that isolated viruses could remain suspended in quiet
air for more than a month, while small droplets of $1\,\mu m$ in size can remain suspended
for several hours, in qualitative agreement with experimental results on virus stability
in aerosols. 
\end{abstract}


\maketitle

{\large \bf Introduction}

With the advance of the Covid-19 virus pandemic a discussion began regarding the proper use of masks as a
safeguard against contagion. The initial advise from the World Health Organization was that only confirmed
infected people should
wear masks, in order to prevent transmission of the virus by sneezing and coughing~\cite{WHO:masksuse}.
But the high rate of
propagation of the virus in different environments around the world raised a concern about this initial
prescription. In particular, the fact that most of infected people go without developing symptoms make them
potential transmitters of the desease. Together with restrictions on human circulation and social
distancing, other important protective measure can be the generalized use of masks in public places,
like hospitals, supermarkets and aircrafts~\cite{Howard2020}.
Although it
is believed that the main mechanism of transmission of the virus is through transport from
infected surfaces to the mouth, nose or eyes, much less is known about transmission through
the air. For how long the virus can remain suspended in the air ? What are dangerous
concentrations ? How an initial concentration is propagated in space and time ? These are
very broad questions, but relevant in order to build a robust response against contagion of
air transmitted deseases, like Covid-19.

The mechanisms of propagation of virus in the air are diverse, depending on the nature
of the process that expels the virus from inside an infected individual. When caughing
or sneezing, a person ejects a turbulent cloud composed of liquid droplets in a wide range
of sizes~\cite{Bourouiba2014,Bourouiba2020}. Larger droplets fall down quickly and the cloud looses
much of its initial
momentum, until a buoyant gas of small droplets of sizes less than $100\, \mu m$ continuous
expanding and spreading through space. Furthermore, in realistic situations, the effects of
air currents due to openings in the builiding or air-conditioning must be taken into account.
Results of simulations of realistic air flows carrying droplets after repeated coughing by a
person in a room show that they can be transported for several meters before they settle
on a surface or fall to the ground, thus having the potential to infect other people around
in times of the order of seconds~\cite{Zhu2006}. It was found that droplets of sizes less than
$\sim 30\,\mu m$
can remain suspended in the air of the room after exiting the main current. Another way of
exhaling the viruses in the air is when speaking and breathing. These are softer processes
as compared with a cough or a sneeze, but they are equally potential ways of spreading the
virus in the surroundings of an infected person. From these observations, it is clear that
the answers to the above posed questions are not unique. Many factors influence the spatial
and temporal extensions of the virus in the air.

Our aim is to present a pedagogical discussion of the motion of the virus in suspended air,
either considering a suspension of isolated viruses or a suspension of small liquid droplets
with a viral load.
When breathing,
sneezing, talking or coughing, droplets with a viral charge are expelled from the mouth of an
infected person. The droplets
dispersed in the sorrounding air do not interact between them.
Small droplets of sizes less than $10\,\mu m$ are
abundant, forming aerosols which can be suspended in the air for a given amount of time.
The main characteristics of the particles which
determine the time of suspension in the air are their mass and size.
In general, there is a distribution of particle masses and sizes in an aerosol. As a consequence of gravity,
heavier particles will fall down at a faster rate than
lighter ones. Nevertheless, recent reports show that the Covid-19 can remain
suspended and active in aerosols for hours~\cite{NEJMc2004973}, making them potentially dangerous for
the transmission of the desease to healthy people breathing around.

We will present some results based on simple physical models for the propagation of the particles with
a viral charge once they come out from the mouth of an infected person. 
First, we will consider a virus that is ejected from the mouth as a free falling object under the only influence
of gravity. We will analise the validity of this approximation with
the size and weight of the particle. When valid, the results for this model predict that the virus should settle down on the ground
after travelling approximately $1$ to $3$ meters, as suggested by different sources and transmitted in the general media
~\cite{WHO:faq}. Then, we will present some estimates of the air interparticle distance as compared with the
size of the Covid-19 virus and of larger droplets. This will suggest us that the virus falls down in a
viscous medium, the air. We will address the problem of the identification of the correct form of the drag
force on the droplets, which has been considered many times in related contexts, see e.g. references
\cite{Purcell1977, Timmerman1999} for excellent discussions (see also ~\cite{Mungan2006,Vial2012} for
elegant analytical solutions to the equations of motion).
Because expelled droplets, where the virus can be found, show a distribution of sizes
and masses, we will present simple estimations of the rate and time of decay for typical sizes of the particles. We will not consider the initial turbulent flow nor
external currents produced by air conditioning, for example~\cite{Bourouiba2014,Zhu2006}. Instead, we will focus
on the late regime of deposition of the small droplets that remain suspended in quiet air
in equilibrium. The results can be considered as bringing lower bounds, or conservative
estimations, to the spatial and temporal extensions of airborne virus.
The analysis leads us to conclude that typical droplets containing virus may remain suspended in quiet air for
more than an hour, in agreement with experimental results reported recently~\cite{NEJMc2004973}. 

\vspace{1cm}
{\large \bf A droplet in free fall}

Free fall is the simplest model for the falling of a particle under the influence of gravity. It is assumed
that there are no collisions with other particles during the trajectory.
The position and velocities of the particles are given by Newton's equations of motion for a free
fall~\cite{Feynman:motion}:
\beqa \label{eq:newton}
x(t) & = & x_0 + v_{0x}t  \\
v_x(t) & = & v_{0x} \nonumber \\
y(t) & = & y_0 + v_{0y}t + \frac{1}{2}g\,t^2 \nonumber \\
v_y(t) & = & v_{0y} + gt, \nonumber
\eeqa
where $x$ and $y$ refer to horizontal and vertical directions respectively, $(x_0,y_0)$ are the initial
horizontal and vertical position coordinates of the particle, $(v_{0x},v_{0y})$ are the corresponding components of
the initial velocity vector, and $g=-9,8\,m/s^2$ is the acceleration of gravity (the minus sign reflects that
distances grow positive from bottom to top, and gravity points downwards in this reference frame).
Here, we will be interested in
knowing how far a particle can travel, once it leaves the mouth of a person with an initial 
velocity vector in the horizontal direction, given by $(v_{0x},v_{0y}=0)$.
The initial position of the particle will be set to $(x_0=0,y_0=h)$, where $h$ is the height at which
the mouth
is located. Within
these conditions, from the third equation from (\ref{eq:newton}) we can obtain the time to reach the floor
($y=0$) from an initial height $y_0=h$:
\beq \label{eq:freefalltime}
t(h) = \sqrt{\frac{2h}{g}}.
\eeq
Then, for example, a particle in free fall from a typical height of a person , $h=1.7\,m$,
will reach the floor in $t(1.7\,m)=0.59\,s$.
The horizontal distance it can travel is then obtained from the first of equations (\ref{eq:newton}). It
depends on the initial horizontal velocity $v_{0x}$.
We chose two representative cases, extracted from an experimental study reported in
\cite{Tang2013}. We considered the maximal reported velocities, a conservative choice.
The maximal initial velocity of a particle when talking or breathing was reported to be $v_1 \approx 1.5\,m/s$,
while the maximal initial velocity when sneezing or caughing was found to be $v_2 \approx 5.0\,m/s$. Although in the cited report
these corresponded to maximal speed, i.e. the scalar value of the velocity vector, we will consider them to
be maximal representative of horizontal velocities, which we assumed in this case. Some typical values of time of
decay and travel distance to the floor are shown in Table \ref{tab:freefall}.

\begin{table}[!]
\begin{tabular}{||c|c|c|c||}
  \multicolumn{4}{c}{}\\ \hline
    $h [m]$       &  \;  $t_h [s]$ \;   &  \;  $x_{v_1}(t_h) [m]$  \;&  \;  $x_{v_2}(t_h) [m]$  \;  \\ \hline
  \; $1.5$  \;    &  0.55               &  0.83                    &   2.77  \\ \hline
    $1.6$         &  0.57               &  0.86                    &   2.86  \\ \hline
 \;  $1.7$    \;  &  0.59               &  0.88                    &   2.95  \\ \hline
  \;  $1.8$    \; &  0.61               &  0.91                    &   3.03  \\ \hline
\end{tabular}
\caption{Times of decay to the floor and horizontal distances of a spittle droplet when talking or breathing ($x_{v_1}$) and after
sneezing or coughing ($v_{v_2}$), for four initial heights computed with a free fall model.}
\label{tab:freefall}
\end{table}

As we can see in Table \ref{tab:freefall}, the times for a particle in free fall to settle on the ground from the
typical height of a person is less than a second. The estimates of the horizontal distance travelled by the virus
is less than $1\,m$ when talking or breathing, and around $3\,m$ from sneezing or coughing. These estimates
roughly agree with others suggesting that keeping a distance of around
$2\,m$ between people could be enough to represent a secure situation~\cite{WHO:faq}. Nevertheless,
as will be shown in the next section,
the free fall approximation is not accurate for the decay of small droplets, which may stay suspended in the air for more
than an hour~\cite{Bar-0n2020,NEJMc2004973}.
In the next section we will analyze the range of validity of the free fall model and show that for particles of sizes less than  $100\,\mu m$ it is necessary to consider friction effects due to the viscosity of air.

\vspace{1cm}
{\large \bf Particle flow in a viscous medium}

The density of dry air at $T=20^oC$
and $1\,atm$ is $\rho_{air}=1.2041\,kg/m^3$~\cite{wiki:airdensity}. Considering the air in a cube of volume $V=l^3$,
we can estimate the typical interparticle distance in air to be:
\beq \label{eq:interdistance}
l \approx (m/\rho_{air})^{1/3},
\eeq
where $m$ is the average mass of an air particle (mainly nitrogen and
oxigen molecules). The average molecular mass of
dry air is $M_{air}=28.97\,g/mol$~\cite{wiki:molarmass}. Considering that there are $N_A=6.022\times 10^{23}$
particles in a mol of
substance, where $N_A$ is Avogadro's constant, then $m =M_{air}/N_A \approx 4,81\times 10^{-26}kg$.
Thus, the typical interparticle
distance in air is $l \approx 3,42\times 10^{-9}\,m =3.42\,nm$
~\footnote{Equivalently, we can define $l = 1/n_a$ , where $n_a$ is the number of particles per unit volume.
  Then, assuming that air at normal pressure and temperature behaves as an ideal gas, $n_a = P/kT$ ,
  where $P$ is the pressure, $T$
the absolute temperature and $k$ the Boltzmann constant. At $T = 293^o K$ and $P = 1\,atm$ one
recovers the result in (\ref{eq:interdistance})}. To compare with, the approximate
size of the Covid-19 virus is $d_C\approx 100\,nm$~\cite{Zhu2019}.
Then, the virus will hit many particles in its trajectory
through the air~\footnote{Another possibility is to compare with the “mean free path”, $l_p = 1/(n \sigma)$,the average distance
a particle will travel between collisions. $n$ is the particle density and $\sigma$ the collisional cross
section. For air at $T = 25^o C$ and normal pressure $P = 1\,atm$, $l_p \approx 34\,nm$. Then, the Covid-19
is still 3 times larger than the mean free path in air.}. As a better model to compute characteristics of the trajectory of the virus  we can
assume that it is an approximately spherical particle falling in a viscous medium, the air.
Furthermore, in the vast majority of cases, the virus will come attached to small liquid droplets expelled when
talking or coughing, which have a larger volume than a single virus.
Then, besides the effective downward gravitational force (the difference between the weight and buoyancy), one
has to consider the upward drag force $F_d$ on the particle. The nature and precise form of the drag force is
a complex issue in fluid dynamics. A well known quantity to characterize the dynamical regime is the
Reynolds number~\cite{Fox1992,Feynman:flow}:
\beq \label{eq:reynolds}
Re = \frac{\rho_f\, d\, v}{\eta},
\eeq
where $\rho_f$ is the fluid density, $d$ is the typical lengthscale of the object (the diameter in the case of a
sphere), $v$ the relative velocity, and $\eta$ is the dynamic viscosity.
For the microscopic objects of relevance for the present problem, the Reynolds number is very small. This claim
will be justified a posteriori, once we compute the typical velocities involved.
At small Reynolds number $Re < 1$, the drag force on a sphere of radius $r$
is given by Stokes' law~\cite{Timmerman1999,Bourouiba2014}:
\beq
F_d=6\pi\,\eta \,r\,v.
\eeq
For sphere diameters less than $1\, \mu m$ a correction factor has to be
considered~\cite{Crowder2002}. As for the sizes of interest in our analysis this factor is of order one, we will
use the approximate expression given above. In the vertical direction,
this upward viscous force grows with time proportional to the velocity. It is opposed by the excess
force between the weight and the buoyancy:
\beq
F_g=(\rho_s-\rho_f)\,g\frac{4}{3}\pi r^3,
\eeq
where $\rho_s$ is the density of the sphere. In this case, we will consider
spherical droplets with $\rho_s=997\,kg/m^3$ (the density of water) and $\rho_f=\rho_{air}$ given above.
As $\rho_s \approx 1000\,\rho_{air}$, we can discard the density of air in $F_g$, which reduceds
to the weight of the spherical droplet, $F_g \approx m_sg$. When the upward drag force $F_d$ equals the
weight of the particle, $F_d=F_g$, the droplet reaches mechanical equilibrium and, from then on,
continues to fall with a constant terminal velocity given by:
\beq \label{eq:termvel}
v_T = \mu \,m_sg,
\eeq
where $\mu=1/6\pi \eta \,r$ is the droplet mobility in the fluid. For air, the dynamical viscosity is
$\eta =18.5\ \mu Pa\cdot s=1.85\times 10^{-5}\,kg \cdot m^{-1}\cdot s^{-1}$. Assuming that particles have an
initial vertical velocity equal to zero, the terminal velocity is the maximal velocity it attains while falling
down~\footnote{In fact, the motion of a particle with a drag force linear in the velocity can be 
  solved in full generality, i.e. for any initial velocity, see e.g. \cite{Timmerman1999,Mungan2006,Vial2012}.
  Nevertheless, for the purposes of this work, it is enough to consider the
particular situation in which the object falls down from rest.}.

First of all, let's make an estimation of the limits of validity of the ``free fall'' model of the previous
section. The
free fall model assumes that gravity, i.e. the weight, is the only relevant force governing the motion of a particle.
Then, as long as the weight is much larger than the drag force, the motion can be considered a free fall.
Now, the weight of a spherical particle $F_g=m_sg$ can be written equivalently as
$F_g=\rho_s V\,g=\rho_s\,(4/3\,\pi r^3)g$. It grows proportional to the volume,
i.e. as the cube of its radius. On the other hand, the drag force produced by the viscosity of air grows
linearly with the radius of the sphere, $F_d=6\pi\,\eta \,r\,v$, much slower than the weight. Then, one can
naively expect that the weight will quickly be much larger than the drag force as the size of the particle
grows. Nevertheless, $F_d$ also
depends on the velocity, which is growing in the initial part of the motion, before it attains its terminal
value. In any case, if the object falls down from rest,
the real velocity when the particle has fallen down a height $h$ will always be smaller than the
corresponding free fall velocity $v_{ff}$, given by the last of equations (\ref{eq:newton}) with the time given by
(\ref{eq:freefalltime}), i.e.
\beq
v_{ff}(h) = \sqrt{2gh}.
\eeq
We can set this value as an upper bound for the velocity of the particle after falling from a height $h$
from rest.
For example, in the present situation considering a typical height $h=1.7\,m$, the free fall velocity is
$v_{ff} = 5.77\,m/s$. Equating the weight with the drag force computed at $v_{ff}$ we can obtain an
upper bound for the radius of the particle $r_c$:
\beq
r_c = \sqrt{\frac{9}{2}\frac{\eta}{\rho_s g} v_{ff}}.
\eeq
When the radius $r \geq r_c$ the drag force becomes irrelevant and the free fall model is valid. For the
case under consideration $r_c \approx 200\,\mu m$. Several studies reported a wide range
of droplet sizes, from the order of $1\,\mu m$ up to millimeters (see. e.g. Table 1 in \cite{Han2013}).
Droplet sizes are strongly
dependent on the nature of the phenomenon, while talking, breathing, sneezing or caughing. For example,
it has been
found that sneezing produces larger droplets than coughing, and also that the velocity at
which the droplets are expelled is considerably larger during sneezing. Other characteristics that may
influence the final results reported in the literature are measurement techniques,
the number of sampled indiviuals and their health conditions, among others.
Last but not least, it is also known that
droplets suffer evaporation after coming out of the respiratory tract. Then, the methodology of measurment
is also important. The further from the mouth, the smaller will be the droplet size, due to evaporation.
For droplets of size $\sim 200\,\mu m$ the time of evaporation is approximately $\approx 6.6\,s$
~\cite{Wells1934}, i.e. larger
than the time to reach the ground
of all cases considered in Table \ref{tab:freefall}. Then, the free fall estimates are relevant for
the motion of sneezing droplets of this size. 

\begin{figure}[ht!]
\centering
\includegraphics[height=.35\textheight,width=.8\textwidth]{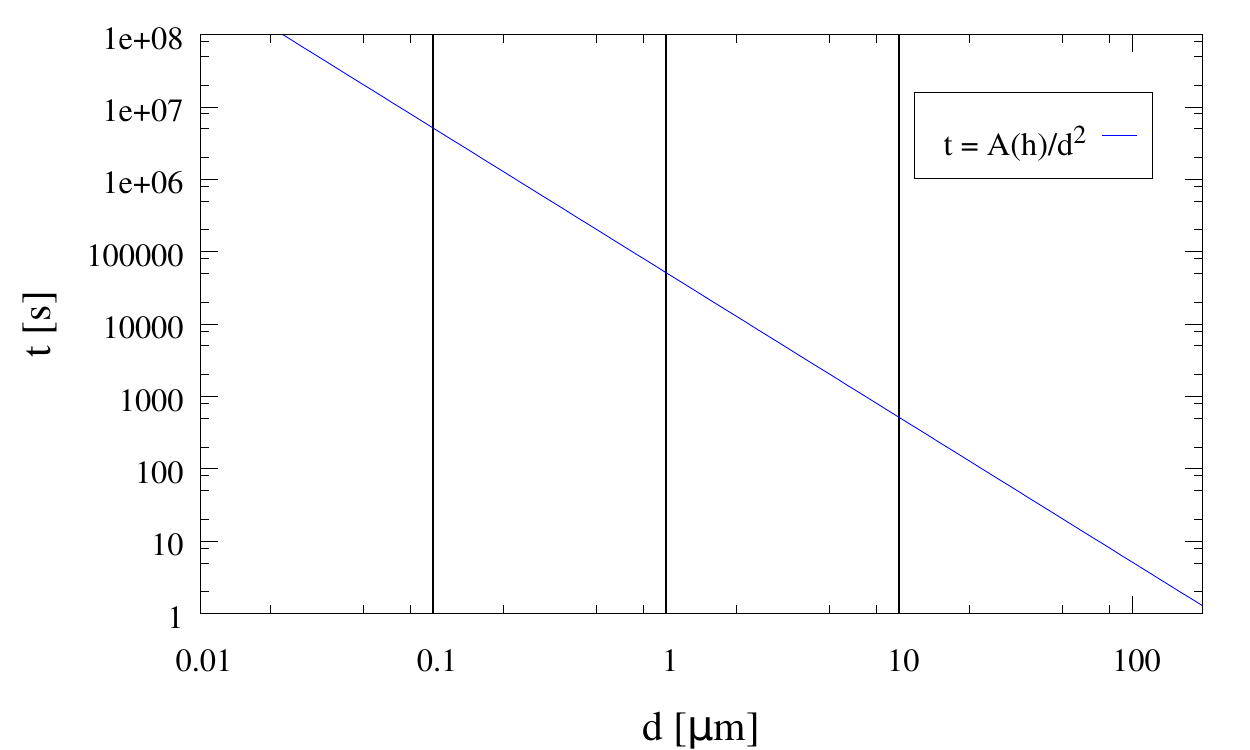}
\caption{Time to reach the floor versus size of droplets emitted while sneezing, coughing, speaking or breathing,
  assuming they fall in viscous air at low Reynolds number. Note the double logarithmic scale. $A(h)$ depends
  on the height from which the droplet is emitted (see text). Vertical solid lines correspond the examples
computed in the text.}
\label{fig:fallingtime}
\end{figure}

Now, let's consider some cases with droplets smaller than $10\,\mu m$, which represent the
majority of cases found in experiments. To begin with,
the smallest droplet of interest is a single Covid-19 virus. The size of the virus (its
approximate diameter) is 
$d_C \approx 100\,nm= 0.1\,\mu m$, and its mass $m_C \approx 10^{-18}\,kg$~\cite{Bar-0n2020}.
Then, its mobility and terminal velocity in dry air will be:
\beqa
\mu & = & \frac{1}{6\pi \eta r} \approx 0.57\times 10^{11}\,s\cdot kg^{-1} \nonumber \\
v_T & = & \mu \,m_Cg \approx 5,59\times 10^{-7}\,m/s .
\eeqa
We note that the downward maximal velocity of a single virus is too small. Assuming that it began its fall with this
vertical velocity  at a height $h = 1.7\,m$, it will reach the ground in a time
$t = h/v_T \approx 3\times 10^6\,s \sim 35\,d$ (see Figure \ref{fig:fallingtime}). Then, single viruses can
remain suspended in undisturbed air for more than a
month before settling down on the ground.

The previous case of single viruses floating freely in the air is probably not the most probable situation,
nor the most dangerous from
a contamination perspective, because the viral charge of the virus depends on its concentration. More relevant
are droplets (mainly composed of water) with a virus charge in them. 
We will consider two typical cases: whole droplets as they come out from the mouth of an infected person,
and droplets nuclei,
that are the remanent of the original droplets once the water has evaporated. 
The typical size of droplet nuclei when coughing is around $d_{dn} \approx 1\,\mu m$ and that of whole droplets
$d_d\approx 10\,\mu m$~\cite{Yang2007}. 
Assuming that the droplets content is mainly water, the terminal velocities and times
of decay from a height $h=1.7\,m$ for both cases are:
\beqa
v_{dn} & \approx & 2.9\times 10^{-5}\,m/s,\hspace{0.5cm} t_{dn}\approx 5.9\times 10^4\,s \approx 16.3\,h \hspace{1cm}
\mbox{for droplets nuclei}, \nonumber \\
v_d & \approx & 2.9\times 10^{-3}\,m/s,\hspace{0.5cm}t_d\approx 5.9\times 10^2\,s \approx 9.8\,mins  \hspace{1cm} \mbox{for water droplets.}
\eeqa
These order of magnitude estimates show that droplets nuclei of $\sim 1\,\mu m$ can remain suspended in air for
around $16$ hours, while droplets of $\sim 10\,\mu m$ can remain for around $10$ minutes. These results are
compatible with recent experimental determinations of the half-life of the Covid-19 virus in
aerosols~\cite{Bar-0n2020,NEJMc2004973}. In Figure \ref{fig:fallingtime} we can see that the sizes
and falling times of relevant droplets emerging from expiratory events span many scales.
From the considerations above, it is an exercise to show that the Stokes'
model used here predicts that the falling time decays with the inverse square of the size:
\beq
t_h(d)=\frac{18\eta}{\rho \,g}\frac{h}{d^2}.
\eeq
For the typical height of a person, which is of interest in the present discussion, 
the effects of $h$ on the final falling time is very small.

Finally, let us consider if the assumption of a drag force linear in the velocity is justified,
i.e. if $Re \leq 1$ for the motions involved in the present problem. Recalling from equation
(\ref{eq:reynolds}) that the Reynolds number is proportional to the size of the object and its
velocity, for the sizes of interest here let's consider a droplet with $d=10\,\mu m$ for which
$v_T=2.9\times 10^{-3}\,m/s$. In this case $Re \approx 1.9\times 10^{-3}$. Then, the problem
considered in this work is in the very small Reynolds number regime. Even for droplets an order
of magnitude larger, $d \sim 100\,\mu m$, the terminal velocity will be of the order of $0.1\,m/s$,
and $Re \sim 1$.

\vspace{1cm}
{\large \bf Diffusion in quiet air}

After realizing that a small droplet can remain suspended in quiet air for several
hours, one can ask how far it can travel before settling on the ground. Assuming that, after
the initial cloud looses most of its initial momentum, the remaining gas undergoes diffusion
in quiet air, we can estimate how far a micro droplet can travel during a given time. In
a diffusive process particles undergo random displacements without a net direction, due to
collisions with other particles. Then, although the average particle displacement is zero, a
measure of how far it goes in its “random walk” in sapce is given by the root mean squared
displacement:
\beq
r_{rms} \equiv \sqrt{\langle r^2(t)\rangle} = (6\,Dt)^{1/2},
\eeq
where $D = \mu k T$ is the diffusion constant~\cite{Feynman:diffusion}, which depends on the mobility of the particle
in the medium and on the temperature. For droplet nuclei of $1\, \mu m$ at $20^o C$, the
diffusion constant is $D \approx 12.13 \times 10^{−12}\, m^2/s$. Then, the order of magnitude of
$r_{rms}$ for a
time span of the order of a few hours is $r_{rms} \sim 1\, mm$. We can conclude that,
in the absence of currents,
micro droplets of these sizes stay practically around the same region where the initial cloud
lost its initial momentum. In a real situation, there will always be some degree of turbulence
in the medium, due to the initial process of sneezing or coughing, or even because
people walk around. Although one cannot expect that diffusion is the
dominant mechanism of propagation of the final cloud of micro droplets, this is useful to
illustrate the slow time scales of dissipation of the gas of droplets in the absence of currents.
Small currents of air are responsible for propagating the remaining cloud of virus-carrying
micro droplets away from the source, creating a potentially infectious atmosphere in closed
spaces which can remain for quite a long time.

\vspace{1cm}
{\large \bf Conclusions}

We have shown estimates of the motion of droplets in the air, depending on droplet size and
initial velocities, as they leave the mouth during talking, breathing, sneezing or coughing. Depending on the
size of the droplets, which can vary in a wide range, two different models can give useful estimates of
the time of persistence of the droplets in quiet air. For droplets larger than $\approx 200\,\mu m$, typical
of sneezing events, a free fall model is appropriate.
They
can attain distances of several meters, justifying recommendations for social distancing.
On the other side, for droplets of sizes less than $\approx 200\,\mu m$, the free fall model is not accurate,
and a model of motion in viscous air is necessary. Our results show that typical respiratory droplets
with sizes $\leq 10\,\mu m$ can remain suspended in dry quiet
air from several minutes to hours. 
The time of
evaporation of water droplets of sizes as small as $10\,\mu m$ is much less than a second~\cite{Wells1934},
leaving droplets nuclei.
Then, we conclude that droplet nuclei, which can be suspended in dry air for hours, may be carriers
of the virus in quiet air environments, like in a closed room or in an aircraft.

Together, these results suggest that airborne transmission of the virus 
may be a possible mechanism of infection. This also points that the use of masks, even for healthy people,
may be a valid way of reducing contagion of Covid-19. More experimental studies are needed in order to
confirm if airborne transmission is relevant as compared with other known mechanisms, like surface mediated ones.

Finally, it is important to stress that the results presented here, based on simplified models of motion,
represent lower bounds
for the time of stability of small droplets in the air. Considering realistic situations, taking
into account small currents and/or convection mechanisms in closed spaces do not contradict the present
conclusions~\cite{Bourouiba2020,Zhu2006,Anchordoqui2020}. Inclusion of these and other effects, like the
process of evaporation of droplets, are interesting routes for future research.

\acknowledgments
I am greateful to Liliana Obreg\'on, Paulo Murilo Castro de Oliveira and Daniel Barci for
useful comments and feedback.
This work was supported in part by a grant from Conselho Nacional de Desenvolvimento
Cient\'ifico e Tecnol\'ogico (CNPq), Brazil.
%
%

\end{document}